\documentclass[11pt,twoside]{article}

\usepackage{asp2006}
\usepackage{graphicx}
\begin{document}
\title{Secular evolution and the assembly of bulges}   
\author{F. Combes} 
\affil{Observatoire de Paris, LERMA, 61 Av. de l'Observatoire, F-75014, Paris}  

\begin{abstract} 
Bulges are of different types, morphologies and kinematics,
from pseudo-bulges, close to disk properties (Sersic index, rotation
fraction, flatenning), to classical de Vaucouleurs bulges,
close to elliptical galaxies.
Secular evolution and bar development can give rise to pseudo-bulges.
To ensure prolonged secular evolution, gas flows are required
along the galaxy life-time. There is growing evidence for cold
gas accretion around spiral galaxies. This can explain
the bar cycle of destruction and reformation, together with
pseudo-bulge formation. However, bulges can also be formed
through major mergers, minor mergers, and massive clumps
early in the galaxy evolution. Bulge formation is so efficient that
it is difficult to explain the presence of bulgeless galaxies today.
\end{abstract}

\section{Secular evolution and bulges, gas flows}  

There are excellent recent reviews on
secular evolution (Kormendy \& Kennicutt 2004, Jogee 2006),
and in particular the formation of bulges has been debated in detail
last year in Oxford, with the IAU Symposium 245 on 
"Formation and Evolution of Bulges".  Only more recent work
will then be reviewed here.

A clear distinction is now well established
between classical bulges and pseudo-bulges, from the 
luminosity distribution (Sersic index, flattening, color)
and the kinematics. On the color-magnitude diagram,
the pseudo-bulges, similar in properties
to disks, are clearly on the blue cloud, while the classical
bulges sit on the red sequence (Drory \& Fisher 2007).
There is a clear bimodality in Sersic index, with the pseudo-bulge
peak at n=1-2, and the classical bulge peak at n=4
 (Fisher \& Drory 2008).

{\bf Gaseous haloes around galaxies}  

 The secular evolution is fueled by external gas accretion,
and there is now growing evidence of gas infalling on
nearby spiral galaxies, although this gas is quite diffuse.
 One of the best example is the edge-on galaxy 
NGC891 (Fraternali et al 2007). HI gas is observed up to
20kpc above the plane, with its rotation decreasing with the altitude. 
Part of this gas could come from galactic fountain, 
but not all, since the angular momentum should then be conserved. 
Moreover, modelisation of the fountain effect predicts gas outflow
(cf the non edge-on galaxy NGC 2403), 
while mostly inflow is observed, like for high velocity clouds in the Milky Way.
Gaseous haloes require accretion of external gas (Fraternali \& Binney 2006).

 A recent review by Sancisi et al (2008)
gathers many examples of extra-planar gas. 
Part of it is due to dwarf companions or
tidal streams, but evidence is mounting
for extragalactic inflow of gas due to cosmic accretion.
This external gas inflow produces
lopsideness and fuels star formation.
When the angular momentum of the accreted gas
is close to perpendicular to that of the galaxy, 
polar rings can form. The simulation
of this phenomenon by Brooks et al (2008) reveals how successively
star formation is occuring in the inner equatorial disk,
then in the outer polar disk. After 1.5 Gyr, the interaction 
between the two disks destroys the polar ring.
The velocity curve is about the same in both
equatorial and polar planes.

{\bf Relative role of gas accretion and mergers}

  In the standard hierarchical scenario, galaxies are thought to assemble
their mass essentially through mergers. However, the gas
accretion from cosmic filaments has been under-estimated.
Analysis of a cosmological simulation
with gas and  star formation shows that most of the 
starbursts are due to smooth flows (Dekel et al 2008).
Corresponding inflow rates are sufficient
to assemble galaxy mass (10-100 M$_\odot$/yr).

\begin{figure*}[t!]
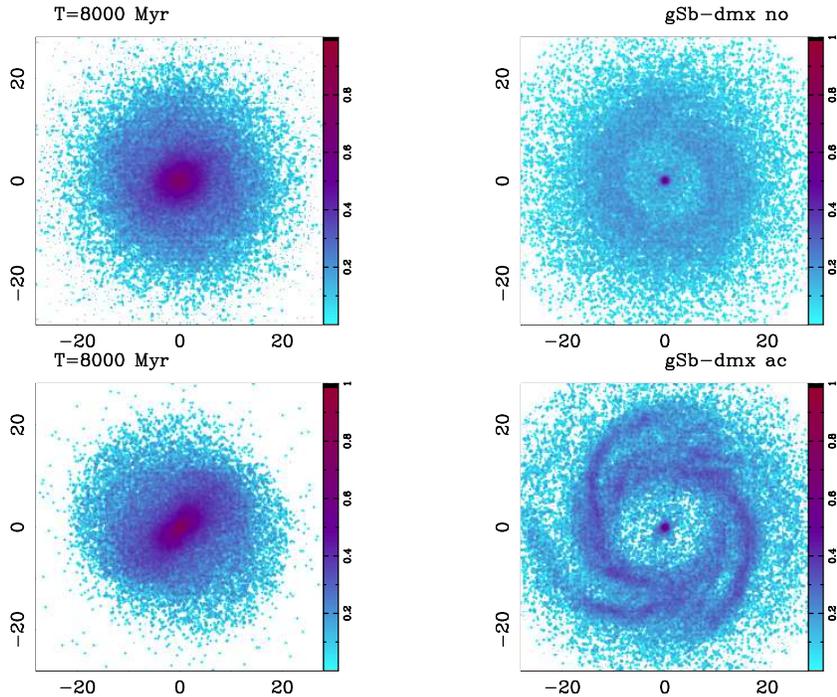

\centering
\includegraphics[clip=true,width=11cm]{combes-f1a.ps}
\includegraphics[clip=true,width=11cm]{combes-f1b.ps}
\caption{Comparison between two Sb galaxy models, without gas accretion
({\bf top}) and with accretion ({\bf bottom}). The Sb model is the maximum
disk model (Combes 2008), the gas accretion rate is 5 M$_\odot$/yr, and the snapshot
corresponds to T=8 Gyr. Note that a bar is maintained only in the
case of gas accretion.
}
\label{fig1}
\end{figure*}

\section{Bar destruction, re-formation, role of gas}

It is now well established that bar redistribution of mass and the vertical resonance
can buiid secularly pseudo-bulges (e.g. Combes et al 1990).
However, in a gaseous disk, not dominated by dark matter,
the bar can quickly weaken and even be destroyed (cf Figure 1).
With only 2\% of the mass, gas infall is enough to transform a bar in a lens
(Friedli 1994, Berentzen et al 1998, Bournaud \& Combes 2002, Bournaud et al 2005).
To pursue bulge formation, and explain the large frequency of bars,
external gas accretion has to be invoked, in a 
self-regulated cycle:
in a first phase, a bar forms through gravitational instability
in a cold disk. The bar produces gas inflow, which  itself
weakens or destroys the bar, through angular momentum transfer to the bar.
Then external gas accretion can replenish the disk, to turn back
to the first phase.

Figure 2 illustrates such cycles, for different galaxy
models, giant disks with different dark matter fractions,
and different dark matter concentrations (Combes 2008).
Although gas is provided at a constant inflow rate in the outer
parts of the disks, it enters the galaxy disk 
by intermittence, and produces starbursts. Indeed,
while the bar is strong, positive torques
between corotation and OLR confine the gas outside OLR.
Only when the bar weakens,
 the gas can replenish the disk, to make
 it unstable again to bar formation. Through
these cycles, the pseudo-bulge can form, as shown
in Figure 3.

\begin{figure}[t!]
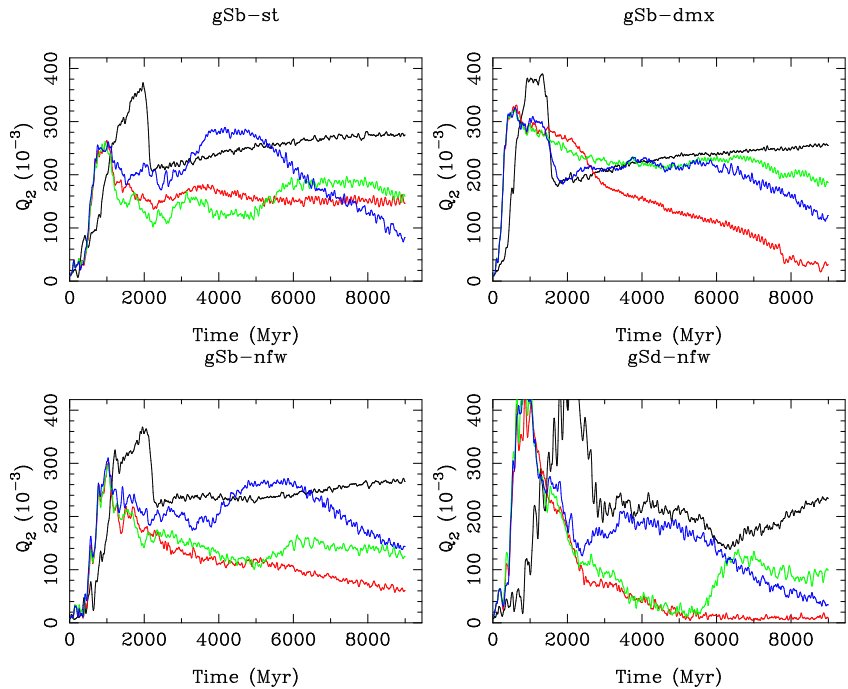

\centering
\includegraphics[clip=true,angle=-90,width=55mm]{combes-f2a.ps}
\includegraphics[clip=true,angle=-90,width=55mm]{combes-f2b.ps}
\includegraphics[clip=true,angle=-90,width=55mm]{combes-f2c.ps}
\includegraphics[clip=true,angle=-90,width=55mm]{combes-f2d.ps}
\caption{Evolution of bar strength, measured as the ratio
Q$_2$ of the maximal $m=2$ tangential force to the radial force, measured
at a radius of 3.3kpc. The black curve corresponds to the purely stellar run,
the red curve, to the spiral galaxy with initial gas, subject to star formation,
but not replenished. These two curves are respectively the top and the bottom
curves. The other curves corresponds to the models with gas accretion,
5 M$_\odot$/yr for the green one, and 5 M$_\odot$/yr for the blue one.
There are 3 galaxy models corresponding to an Sb galaxy, with
different mass ratios between disk and halo, and an Sd galaxy,
with an NFW dark profile.
}
\label{fig2}
\end{figure}

\bigskip
{\bf Formation in a cosmological context}

In cosmological simulations, it is possible to take
into account cosmic gas accretion more realistically.
When spatial resolution is sufficient, it is possible to
see bars form, destroy and reform (Heller et al 2007).

There is clearly in all these
processes the influence of the adopted gas physics:
bar formation is more easy with isothermal gas,
while adiabatic gas heats and prevents disk instability.
Star formation and feedback are then important keys
to regulate the dynamics.

\begin{figure*}[t!]
\centering
\includegraphics[clip=true,width=55mm]{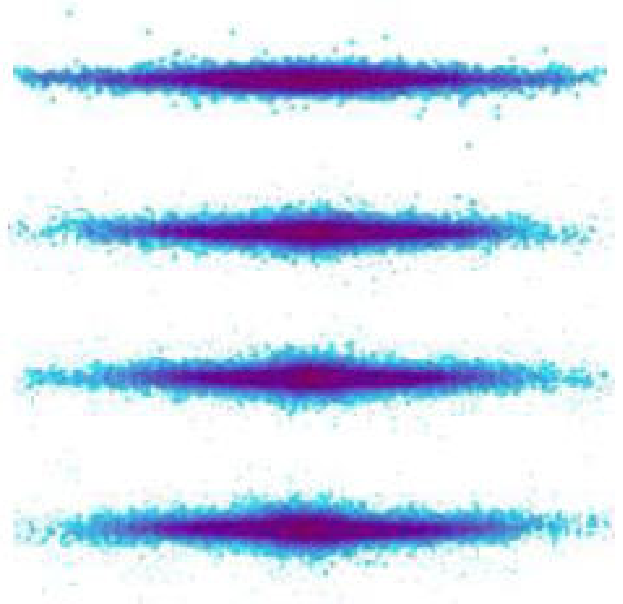}
\includegraphics[clip=true,width=55mm]{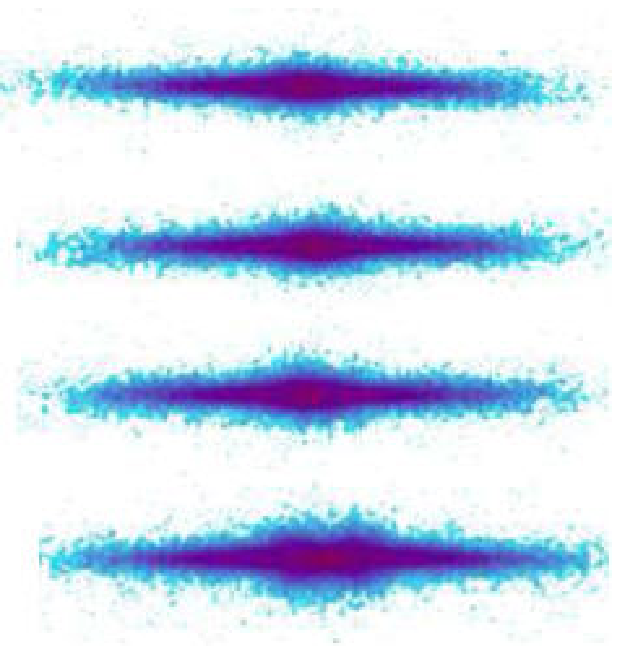}
\caption{Formation of a peanut bulge, during bar formation,
destruction, and reformation, for the standard Sb model. Time is running
from top left to bottom left, then from top right to bottom right.
}
\label{fig3}
\end{figure*}

{\bf Angular momentum transfer with dark matter haloes}

In presence of a massive dark halo, the angular momentum
transfer is essentially towards the dark component, which
helps to reform the bar. 
However, the bar destroys more quickly in presence of gas
(Berentzen et al 2007).
The weakening of the bar is then interpreted as due
to the central mass concentration provided by the gas.
It cannot be due to the vertical resonance, which 
is also weakened or suppressed in presence of large
quantities of gas.

\section{Bar and bulge statistics and high z evolution}

The frequency of bars has been quantified on many samples,
and in particular in the near infrared, where bars are easier
to define (e.g. the OSU NIR sample, Eskrige et al 2002).
 In all samples, the main result is the
paucity of weak bars (Marinova \& Jogee 2007).
Whatever the tool to measure bar strength, the number of strong
bars is high at z=0 (Whyte et al 2002, Block et al 2002, Buta et al 2004).

{\bf Bar frequency with redshift}

If at z=0 about 2/3 of galaxies are barred, with at
least 30\% strongly barred, the strong bars (ellipticity higher
than 0.4) in the optical remain about 30\% at
redshift between 0.2 and 1 (Jogee et al 2004).
At high z, however, results are more uncertain, because of the K-correction,
and the lack of spatial resolution, preventing to detect small bars.
From a recent study of the COSMOS field, the bar
fraction is found to decrease at high z (Sheth et al 2008).
This is easy to interprete in terms of gas fraction: galaxy disks
are more gas rich at high redshifts, and the gas inflow is destroying bars. 
The time spent in a barred phase is then expected to be smaller for
galaxies at z $\sim$ 1.

{\bf B/T and $n$ statistics}

The bulge-to-total luminosity ratio (B/T) and the Sersic index $n$
have been studied by several groups in near-infrared samples.
A clear decrease of B/T and $n$ has been found
with the Hubble type, whatever the barred
or unbarred character (Laurikainen et al 2007).
In the OSU sample of 146 bright spirals in H-band,
where 2/3 of galaxies are barred, 
 60\% have $n<2$, and B/T $<$ 0.2,  barred or not
  (Weinzirl, Jogee, Khochfar et al 2008).
There is in addition a clear correlation between B/T and $n$.

This large observed fraction of low bulge galaxies is
a constraint for models.
In $\Lambda$CDM, a B/T$<$0.2 galaxy requires no merger since 10 Gyr
(or last merger before $z>2$).
 The predicted fraction of these low-bulge bright spiral is 15 times lower
than observed (Weinzirl  et al 2008).
Most of these low-bulge bright spirals must be explained either by 
rare minor mergers or secular evolution, without mergers.
With semi-analytical criteria,
Koda et al (2007) propose a solution in terms of the tail of the 
distribution. 

{\bf Frequency of bulge-less galaxies}

Locally, about 2/3 or the bright spirals are bulgeless, or with
a low-bulge (Kormendy \& Fisher 2008, Weinzirl et al 2008).
Some of the remaining have both a classical bulge and a pseudo-bulge,
plus nuclear clusters (B\"oker et al 2002).

From the observed frequency of edge-on superthin galaxies,
 Kautsch et al (2006) estimate that
1/3 of galaxies are completely bulgeless.
In the SDSS sample, 20\% of bright spirals are bulgeless until z=0.03 
(Barazza et al 2008).
Disk-dominated galaxies are more barred than bulge-dominated ones.
How can this be reconciled with the hierarchical scenario?

\begin{figure*}[t!]
\centering
\includegraphics[clip=true,angle=-90,width=11cm]{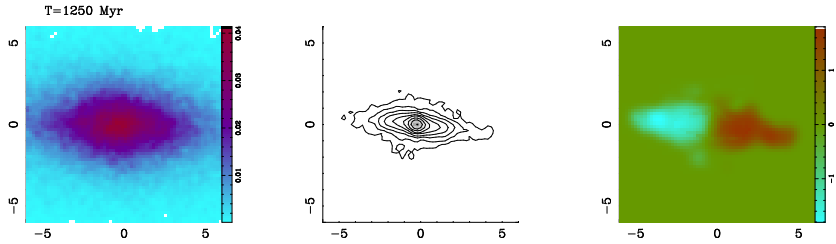}
\caption{Result of the major merger of two equal-mass spiral galaxies,
after 1250 Myr. Left is the stellar density, middle the gas density contours, and right
the gas velocity field, where the wedge indicate the amplitude of the projected
velocity in units of 100km/s. Not all mergers end up in an 
elliptical galaxy, but a significant pseudo-bulge is formed (Crocker et al 2008).
}
\label{fig4}
\end{figure*}

\section{Mergers and bulge formation scenarios}
We can first remark that major mergers do not 
always lead to spheroids, or classical bulges,
but sometimes also to pseudo-bulges.
An extreme example is shown in Figure 4, the
result of an equal mass merger, to reproduce the NGC 4550 system.
 This simulation is run with one disk prograde with
respect to the relative orbit, and one disk
retrograde, which leads to counter-rotation
(e.g. Di Matteo et al 2007).
In this system, the gas eventually
settles in the prograde sense, in corotation with
the prograde stellar disk, which is also the most perturbed
in the interaction, thus ends up as a thicker disk.
The angular momentum exchange between stars, gas and the 
orbit can be seen in Figure 5 (Crocker et al 2008).
This major merger forms a bulge with low $n \sim 1-2$.
This must also be the case for similar encounter geometries, 
i.e. for almost aligned or anti-aligned spins.

\begin{figure*}[t!]
\centering
\includegraphics[clip=true,angle=-90,width=11cm]{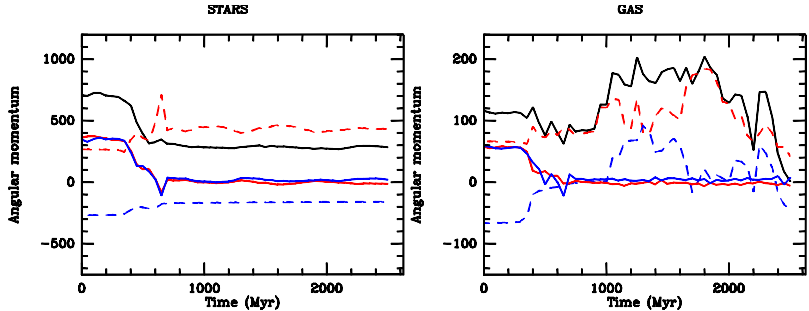}
\caption{Angular momentum evolution of the stars (left) and
gas (right) in the major merger simulation of Fig 4. The red curves
correspond to the prograde galaxy, and the blue curves to the retrograde.
The solid lines indicate the orbital angular momentum,
while the dash lines indicate the internal spins. The scale is in units
of 2.3 10$^{11}$ M$_\odot$ kpc km/s (Crocker et al 2008).
}
\label{fig5}
\end{figure*}

{\bf Scenarios of bulge formation}

Although four different processes have been identified
to form bulges,
major mergers, minor mergers, bars and clumpy young
galaxies, it is usually not easy to separate them
in each galaxy, since they can occur successively.

In major mergers, the tidal trigger first forms strong bars in 
the partner galaxies, which drive the gas inward;
this forms first a pseudobulge in each galaxy.
Then the merger of the two galaxies could provide a classical bulge,
according to the encounter geometry,
which will co-exist with the pseudo ones.

Alternatively, after a classical bulge has formed in a system,
subsequent gas accretion could re-form a disk and a bar, which 
drives the gas towards the center, and form a pseudobulge.

It is likely that most galaxy disks begin gas-rich,
without central concentration, without bulge, and therefore
are highly unstable to form clumpy galaxies at high z. 
Simulations show that there is rapid formation of 
an exponential disk and a bulge, through dynamical friction
(Noguchi 1999, Bournaud et al 2007b). The
evolution is slightly quicker than with spirals and bars.
The rapid bulge formation is
again a problem for the bulgeless galaxies today.

\bigskip
{\bf  Clues from high z galaxies}

Spheroids appear in place quite early (Conselice 2007).
There is a deficit of disk galaxies at z=1.
Could it be a bias of the observations with limited
sensitivity?
Or disky galaxies have formed only recently, 
and in poor environment?
Big disks in rotation are however observed
(Genzel et al 2008, Neichel et al 2008).

Massive bulges (B/T $>$0.2) and ellipticals have the
same early formation,  as shown by
the GOODS study of 0.1 $<$z $<$1.2 galaxies (MacArthur et al 2008).
Their star formation history (SFH) is compatible with a single 
early burst. 
There is however a degeneracy, the same SFH can be obtained,
if the mass is assembled more recently
from dry mergers.

{\bf Multiple minor mergers}

The bulges formed by minor mergers are often the same
as for major mergers, since they are more numerous.
The issue is not the mass ratio of individual mergers, 
but the total mass accreted.  As soon as a given
spiral galaxy has accreted 30-40\% of its initial mass,
either in one event, or a series of small minor mergers,
then the final result is likely to be an elliptical galaxy.
Bournaud et al (2007a) have shown through simulations that 
50 mergers of 50:1 mass ratio can easily 
 form an elliptical, and this is certainly
more frequent than a 1:1 merger.

\section{Conclusion}

Secular evolution and bars play a major role in pseudo-bulge formation,
and this explains their presence in the blue sequence of galaxies.
Since gas inflow weakens or destroy the bar, a galaxy can have several
bar episodes, and each accumulates mass in the pseudo-bulge.
Cold gas accretion, from cosmic filaments, can replenish galaxy disks,
and reform bars. It is expected that
bars were destroyed more frequently at high z, since galaxy disks 
contained more gas. This is in line with the observation of
decreasing bar fraction at high z.

There are several scenarios for bulge formation, classical bulges
through major mergers, but also a succession of minor mergers, 
clumps in young galaxies, coalescing in the center, due to dynamical
friction,  bars.. In most galaxies, there is coexistence of many processes
and it is not easy to reconstruct the dynamical history of
the bulge assembly.
In any case, it is quite easy to form a bulge, in any galaxy environment,
and it is a surprise to observe a large
fraction of bulgeless galaxies. It is 
a challenge both for the hierarchical scenario, but also for
the secular evolution.
It is difficult in particular to find high-z precursors
of these bulgeless galaxies. Have those disks formed recently?

\acknowledgements 
Many thanks to Shardha Jogee and the organising committee 
to invite me at this stimulating conference.

\end{document}